\documentstyle[12pt]{article}

\makeatletter
\@addtoreset{equation}{section}
\makeatother

\def\theequation{\thesection.\arabic{equation}}

\begin{document}

\begin{titlepage}

\begin{flushright}
FIAN/TD/14/03\\
September 03\\
\end{flushright}

\vspace{3cm}

\begin{center}
{\Large\bf On the transformations of hamiltonian gauge algebra}
\end{center}

\begin{center}
{\Large\bf under rotations of constraints}
\end{center}

\vspace{1cm}

\begin{center}
{\large Igor Batalin$^{\,\dag}$ and Igor Tyutin$^{\,\ddag}$ }
\end{center}
\begin{center}
{\em Lebedev Physical Institute RAS,\\
53, Leninsky Prospect, 119991 Moscow, Russia}
\end{center}

\vspace{1.7cm}

\begin{abstract}
By explicit calculation of the effect of a ghost-dependent
canonical transformation of BRST-charge, we derive the
corresponding transformation law for structure coefficients of
hamiltonian gauge algebra under rotation of constraints.We show
the transformation law to deviate from the behaviour (expected
naively) characteristic to a genuine connection.
\end{abstract}

\begin{quote}

\vfill \hrule width 5.cm \vskip 2.mm $^\dag$ {\small \noindent
E-mail: batalin@lpi.ru}

$^\ddag$ {\small \noindent E-mail: tyutin@lpi.ru}

\end{quote}

\end{titlepage}

\newpage

\section{Introduction}

In a hamiltonian constrained (classical) theory, as physical dynamics lives on
the constraint surface
\begin{equation}
T_{\alpha}(q,p)=0,
\end{equation}
the constraint functions $T_{\alpha}(q,p)$ themselves are only determined up
to arbitrary rotation
\begin{equation}\label{0.2}
T_{\alpha}\quad\longrightarrow\quad T^{\prime}_{\alpha}=\Lambda^{\beta
}_{\alpha}T_{\beta},
\end{equation}
with an invertible matrix $\Lambda^{\beta}_{\alpha}(q,p)$.

In the case of being $T_{\alpha}(q,p)$ first-class constraints \cite{1}, which
we restrict ourselves to in what follows, the Poisson bracket involution
relations hold,
\begin{equation}
\{T_{\alpha},T_{\beta}\}=U_{\alpha\beta}^{\gamma}T_{\gamma},\label{0.3}%
\end{equation}
with some structure coefficients $U_{\alpha\beta}^{\gamma}(q,p)$ antisymmetric
in $(\alpha,\beta)$. These coefficients are only determined up to arbitrary
trivial shift
\begin{equation}
U_{\alpha\beta}^{\gamma}\quad\longrightarrow\quad U_{\alpha\beta}^{\gamma
}+{\cal A}_{\alpha\beta}^{\delta\gamma}T_{\delta},\label{0.4}%
\end{equation}
with any ${\cal A}_{\alpha\beta}^{\delta\gamma}(q,p)$ antisymmetric in
$(\alpha,\beta)$ and $(\delta,\gamma)$.

As the involution relations (\ref{0.3}) should be covariant w.r.t. rotations
(\ref{0.2}), the structure coefficients $U^{\gamma}_{\alpha\beta}$ should be
transformed like a sort of connection,%

\begin{eqnarray}
& U_{\alpha\beta}^{\gamma}\quad\longrightarrow\quad U^{\prime}{}_{\alpha\beta
}^{\gamma}=\Lambda_{\alpha}^{\mu}\Lambda_{\beta}^{\nu}U_{\mu\nu}^{\rho}
\Lambda_{\rho}^{-1\gamma}+\Lambda_{\lbrack\alpha}^{\mu}\{T_{\mu}%
,\Lambda_{\beta]}^{\nu}\}\Lambda_{\nu}^{-1\gamma}+\nonumber\\
& +\frac{1}{2}\{\Lambda_{\alpha}^{\{\mu},\Lambda_{\beta}^{\nu\}}\}T_{\mu
}\Lambda_{\nu}^{-1\gamma}+A_{\alpha\beta}^{\mu\nu}T_{\mu}\Lambda_{\nu
}^{-1\gamma},\nonumber\\
& {\cal A}_{\alpha\beta}^{\prime\delta\gamma}=A_{\alpha\beta}^{\mu\nu
}\Lambda_{\mu}^{-1\delta}\Lambda_{\nu}^{-1\gamma}.\label{0.5}%
\end{eqnarray}
The fourth term in r.h.s. of (\ref{0.5}) is not actually required by the
covariance of (\ref{0.3}), but it appears, in general, to represent an
admixture of the arbitrariness (\ref{0.4}) in the transformation (\ref{0.5}).

It is well-known \cite{FraVil75} - \cite{FraFra78} that the involution
relations (\ref{0.3}) are generated, together with all their compatibility
conditions, by means of hamiltonian master equation,
\begin{equation}
\{\Omega,\Omega\}=0,\quad\varepsilon(\Omega)=1,\quad\mathrm{gh}(\Omega
)=1,\label{0.6}%
\end{equation}
when expanded in power series in ghost coordinates $C^{\alpha}$ and momenta
$\bar{{\cal P}}_{\alpha}$, gh($C^\alpha$)$=-$gh($\bar{\cal P}_\alpha$)=1, as
follows
\begin{equation}
\Omega=C^{\alpha}T_{\alpha}+\frac{1}{2}C^{\beta}C^{\alpha}U_{\alpha\beta
}^{\gamma}\bar{{\cal P}}_{\gamma}+\ldots,\label{0.7}%
\end{equation}
where dots mean higher-order terms in $C^{\alpha}$, $\bar{{\cal P}}%
_{\alpha}$.

Master equation (\ref{0.6}) allows for ghost-dependent canonical
transformations,
\begin{equation}
\Omega\quad\longrightarrow\quad\Omega^{\prime}=U\Omega,\;U=e^{\mathrm{ad}%
G}\quad\varepsilon(G)=0,\quad\mathrm{gh}(G)=0,\label{0.8}%
\end{equation}
where $(\mathrm{ad}X)Y\equiv\{X,Y\}$, and
\begin{equation}
G=G_{0}+C^{\alpha}G_{1\alpha}^{\beta}\bar{{\cal P}}_{\beta}+\ldots
,\label{0.9}%
\end{equation}
with $G_{0}(q,p)$, $G_{1\alpha}^{\beta}(q,p)$, $\ldots$, being arbitrary
coefficient functions of a generators $G$. It is obvious that $\exp
(\mathrm{ad}G_{0})$ itself just corresponds to arbitrary canonical
transformation in original phase space, while
\begin{equation}
\Lambda_{\alpha}^{\beta}=\left(  e^{G_{1}}\right)  _{\alpha}^{\beta
}\label{0.10}%
\end{equation}
describe rotations (\ref{0.2}).

In the present paper we are going to show by explicit calculation that the
transformations (\ref{0.8}) yield $A^{\mu\nu}_{\alpha\beta}\neq0$ in
(\ref{0.5}) even if $G^{\beta}_{1\alpha}$ are the only nonzero coefficients in
(\ref{0.9}), i.e. $G=C^{\alpha}G^{\beta}_{1\alpha}\bar{{\cal P}}_{\beta}$.

On the other hand, we will show that the trivial shift part induced in
(\ref{0.5}) by pure rotations (\ref{0.2}) can always be compensated by the
effect of $CC\bar{{\cal P}}\bar{{\cal P}}$-order contribution, chosen
appropriately, to the generator(\ref{0.9}).

As usual, $\varepsilon(A)$ and gh($A$) denote, respectively, the Grassmann
parity and ghost number of a quantity $A$, while $\{A,B\}$ and $[A,B]$ stand,
respectively, for Poisson superbracket and supercommutator for any quantities
$A$ and $B$.

For the sake of simplicity, we restrict ourselves to the case of purely
bosonic original phase variables $q,p$, so that all constraints $T_{\alpha}$
are bosonic as well, while all ghosts $C^{\alpha}$, $\bar{{\cal P}}%
_{\alpha}$ are fermionic.

\section{Classical transformations}

Let us represent the $\lambda$-parameterized canonical
transformation (\ref{0.8}) by means of an ordinary differential
equation of the form
\begin{equation}
\frac{d\Omega^{\prime}}{d\lambda}=\{G(\lambda),\Omega^{\prime}\},\quad\left.
\Omega^{\prime}\right|  _{\lambda=0}=\Omega, \label{1.1}%
\end{equation}
\begin{equation}
\Omega^{\prime}=U\Omega,\quad
U=P_{\lambda}e^{\int_{0}^{\lambda}d\lambda^{\prime}\mathrm{ad}G
(\lambda^{\prime})}=e^{\mathrm{ad}\tilde{G}(\lambda)},\label{1.1a}%
\end{equation}

where generator $G$ is given by (\ref{0.9}), and
\begin{equation}
\Omega^{\prime}=C^{\alpha}T_{\alpha}^{\prime}(\lambda)+\frac{1}{2}C^{\beta
}C^{\alpha}U_{\alpha\beta}^{\prime\gamma}(\lambda)\bar{{\cal P}}_{\gamma
}+\ldots.\label{1.2}%
\end{equation}

In the case of a pure rotation of constraints we have
\begin{equation}
G=C^{\alpha}G_{1\alpha}^{\beta}\bar{{\cal P}}_{\beta}\label{1.3}%
\end{equation}
(we assume that $G_{1\alpha }^{\beta }$ does not depend on $\lambda $), which
implies
\begin{equation}
\frac{dT^{\prime}_\alpha}{d\lambda}=G_{1\alpha}^{\beta}T_{\beta}^{\prime}%
,\quad\left.  T_{\alpha}^{\prime}\right|  _{\lambda=0}=T_{\alpha},\label{1.4}%
\end{equation}
and
\begin{equation}
\frac{dU_{\alpha\beta}^{\prime\gamma}}{d\lambda}=-G_{1[\alpha}^{\delta
}U_{\beta]\delta}^{\prime\gamma}-U_{\alpha\beta}^{\prime\delta}G_{1\delta
}^{\gamma}+\{T_{[\alpha}^{\prime},G_{1\beta]}^{\gamma}\},\quad\left.
U_{\alpha\beta}^{\prime\gamma}\right|  _{\lambda=0}=U_{\alpha\beta}^{\gamma
}.\label{1.5}%
\end{equation}
Eq. (\ref{1.4}) yields immediately
\begin{equation}
T_{\alpha}^{\prime}=\left(  e^{\lambda G_{1}}\right)  _{\alpha}^{\beta
}T_{\beta}.\label{1.6}%
\end{equation}
Let us seek for a solution to (\ref{1.5}) in the form
\begin{equation}
U_{\alpha\beta}^{\prime\gamma}=\left(  e^{\lambda G_{1}}\right)  _{\alpha
}^{\mu}\left(  e^{\lambda G_{1}}\right)  _{\beta}^{\nu}\bar{U}_{\mu\nu}^{\rho
}\left(  e^{-\lambda G_{1}}\right)  _{\rho}^{\gamma}.\label{1.7}%
\end{equation}
It follows from (\ref{1.5}) that
\begin{equation}
\left(  e^{\lambda G_{1}}\right)  _{\alpha}^{\mu}\left(  e^{\lambda G_{1}%
}\right)  _{\beta}^{\nu}\frac{d\bar{U}_{\mu\nu}^{\rho}}{d\lambda}\left(
e^{-\lambda G_{1}}\right)  _{\rho}^{\gamma}=\{T_{[\alpha}^{\prime},G_{1\beta
]}^{\gamma}\},\quad\left.  \bar{U}_{\mu\nu}^{\rho}\right|  _{\lambda=0}%
=U_{\mu\nu}^{\rho},\label{1.8}%
\end{equation}
and, hence,
\begin{eqnarray}
U_{\alpha\beta}^{\prime\gamma}=\left(e^{\lambda G_{1}}\right)_\alpha^\mu
\left(e^{\lambda G_{1}}\right)_{\beta}^{\nu}U_{\mu\nu}^{\rho}
\left(e^{-\lambda G_{1}}\right)_{\rho}^{\gamma}+ \nonumber \\
+\int_0^\lambda d\lambda^\prime
\left(e^{(\lambda-\lambda^{\prime})G_1}\right)_\alpha^\mu
\left(e^{(\lambda-\lambda^{\prime})G_1}\right)_\beta^\nu
\left\{T_{[\mu}^{\prime}(\lambda^{\prime}),G_{1\nu]}^\rho\right\}
\left(e^{-(\lambda-\lambda^{\prime})G_1}\right)_\rho^\gamma. \label{1.9}
\end{eqnarray}
Then, by making use of the variation formula
\begin{equation}
\delta e^{\lambda G_{1}}=\int_{0}^{\lambda}d\lambda^{\prime}e^{\lambda
^{\prime}G_{1}}\delta G_{1}e^{(\lambda-\lambda^{\prime})G_{1}},\label{1.10}%
\end{equation}
one can rewrite (\ref{1.9}) in the form
\begin{eqnarray}
U_{\alpha\beta}^{\prime\gamma}=\left(e^{\lambda G_1}\right)_\alpha^\mu
\left(e^{\lambda G_1}\right)_\beta^\nu U_{\mu\nu}^\rho
\left(e^{-\lambda G_1}\right)_\rho^\gamma+ \nonumber \\
+\left(e^{\lambda G_1}\right)_{[\alpha}^\mu\left\{T_\mu,
\left(e^{\lambda G_1}\right)_{\beta]}^{\nu}\right\}
\left(  e^{-\lambda G_{1}}\right)_{\nu}^{\gamma }+\nonumber\\
+\frac{1}{2}\left\{  \left(  e^{\lambda
G_{1}}\right)  _{\alpha}^{\{\mu },\left(  e^{\lambda G_{1}}\right)
_{\beta}^{\nu\}}\right\}  T_{\mu}\left( e^{-\lambda G_{1}}\right)
_{\nu}^{\gamma}+A_{\alpha\beta}^{\mu\nu}T_{\mu }\left(  e^{-\lambda
G_{1}}\right)  _{\nu}^{\gamma},\label{1.11}
\end{eqnarray}
where
\[
A_{\alpha\beta}^{\mu\nu}=\frac{1}{2}\int_{0}^{\lambda}d\lambda^{\prime}\left(
e^{(\lambda-\lambda^{\prime})G_{1}}\right)  _{\alpha}^{\rho}\left(
e^{(\lambda-\lambda^{\prime})G_{1}}\right)  _{\beta}^{\sigma}\left\{  \left(
e^{\lambda^{\prime}G_{1}}\right)  _{[\rho}^{[\mu},G_{1\sigma]}^{\tau}\right\}
\left(  e^{\lambda^{\prime}G_{1}}\right)  _{\tau}^{\nu]}+
\]
\begin{equation}
+A_{\alpha\beta}^{\mu\nu\delta}T_{\delta},\label{1.12}
\end{equation}
and $A_{\alpha\beta}^{\mu\nu\delta}$ are arbitrary functions
totally antisymmetric in ($\alpha$, $\beta$) and ($\mu$, $\nu$, $\delta$).

Solution (\ref{1.11}) has the structure (\ref{0.5}) with $\Lambda_{\beta
}^{\alpha}=\left(  e^{\lambda G_{1}}\right)  _{\beta}^{\alpha}$ and
$A_{\alpha\beta}^{\mu\nu}$ defined by (\ref{1.12}) in terms of $G_{1\beta
}^{\alpha}$.

Let us represent $A_{\alpha\beta}^{\mu\nu}$ (\ref{1.12}) up to a few low
orders in $G_{1\beta}^{\alpha}$. In the zero, first and second order, we have
\begin{equation}
\left(  A_{\alpha\beta}^{\mu\nu}\right)  _{0}=0,\;\left(  A_{\alpha\beta}%
^{\mu\nu}\right)  _{1}=0,\;\left(  A_{\alpha\beta}^{\mu\nu}\right)
_{2}=0,\label{1.13}%
\end{equation}
while the third-order contribution reads
\begin{equation}
\left(  A_{\alpha\beta}^{\mu\nu}\right)  _{3}=\lambda^{3}\left[  \frac{1}%
{12}\left\{  \left(  G_{1}^{2}\right)  _{[\alpha}^{[\mu},G_{1\beta]}^{\nu
]}\right\}  +\frac{1}{6}\left\{  G_{1[\alpha}^{[\mu},G_{1\beta]}^{\rho
}\right\}  G_{1\rho}^{\nu]}\right]  \not \equiv O(T),\label{1.14}%
\end{equation}
see Appendix for the detail.

Now, let us modify the generator (\ref{1.3}) by adding
$CC\bar{\cal P}\bar{\cal P}$- and
$CCC\bar{\cal P}\bar{\cal P}\bar{\cal P}$-terms,
\begin{equation}
G=C^{\alpha }G_{1\alpha }^{\beta }\bar{\cal{P}}_{\beta }+\frac{1}{4}
C^{\beta }C^{\alpha }G_{2\alpha \beta }^{\gamma \delta }(\lambda )
\bar{\cal P}_{\delta }\bar{\cal P}_{\gamma }+\frac{1}{36}C^\gamma
C^{\beta }C^{\alpha }G_{3\alpha \beta \gamma }^{\delta \rho \sigma}
(\lambda )\bar{\cal P}_{\sigma}\bar{\cal P}_\rho\bar{\cal P}_\delta,
\label{1.15}
\end{equation}
where new coefficient functions $G_{2\alpha \beta }^{\gamma \delta }$ are
antisymmetric in ($\alpha $, $\beta $) and ($\gamma $, $\delta $), while
$G_{3\alpha \beta \gamma }^{\delta \rho \sigma }$ are antisymmetric in
($\alpha $, $\beta $, $\gamma $) and ($\delta $, $\rho $, $\sigma $).

Then, Eq. (\ref{1.4}) remains the same, while the only modification in Eq.
(\ref{1.5}) is that the inhomogenity in its r.h.s. acquires a new term,
$G_{2\alpha\beta}^{\mu\gamma}T_{\mu}^{\prime}$, so that the solution
(\ref{1.11}) for $U_{\alpha\beta}^{\prime\gamma}$ remains valid with the new
generator (\ref{1.15}) if one modifies the expression for $A_{\alpha\beta
}^{\mu\nu}$ as follows
\[
A_{\alpha\beta}^{\mu\nu}=\int_{0}^{\lambda}d\lambda^{\prime}\left(
e^{(\lambda-\lambda^{\prime})G_{1}}\right)  _{\alpha}^{\rho}\left(
e^{(\lambda-\lambda^{\prime})G_{1}}\right)  _{\beta}^{\sigma}\times
\]
\begin{equation}
 \times\left[  \frac{1}{2}\left\{  \left(  e^{\lambda^{\prime}G_{1}}\right)
_{[\rho}^{[\mu},G_{1\sigma]}^{\tau}\right\}  \left(  e^{\lambda^{\prime}G_{1}%
}\right)  _{\tau}^{\nu]}+G_{2\rho\sigma}^{\tau\kappa}\left(
e^{\lambda^{\prime}G_{1}}\right)  _{\tau}^{\mu}\left(  e^{\lambda^{\prime
}G_{1}}\right)  _{\kappa}^{\nu}\right]  +A_{\alpha\beta}^{\mu\nu\delta
}T_{\delta}\label{1.16}%
\end{equation}
Note that the $CC\bar{\cal P}\bar{\cal P}$-term in (\ref{1.15}) is the
highest-order one to contribute to the transformation law for $U_{\alpha
\beta }^{\gamma }$.

Then, if one chooses
\begin{equation}
G_{2\rho\sigma}^{\tau\kappa}=\frac{1}{2}\left\{  \left(  e^{\lambda G_{1}%
}\right)  _{[\rho}^{\zeta},G_{1\sigma]}^{[\tau}\right\}  \left(  e^{-\lambda
G_{1}}\right)  _{\zeta}^{\kappa]}\label{1.17}%
\end{equation}
(and $A_{\alpha\beta}^{\mu\nu\delta}=0$), then the expression (\ref{1.16})
vanishes.

In principle, the above calculations can be extended to higher structure
coefficient functions in (\ref{1.2}). As an example, we consider the next,
$CCC\bar{\cal P}\bar{\cal P}$-order contribution to (\ref{0.7}),
\begin{equation}
\Omega=C^{\alpha}T_{\alpha}+\frac{1}{2}C^{\beta}C^{\alpha}U_{\alpha\beta
}^{\gamma}\bar{{\cal P}}_{\gamma}+\frac{1}{12}C^\gamma C^{\beta}C^{\alpha
}U_{\alpha\beta\gamma}^{\delta\rho}\bar{{\cal P}}_{\rho}\bar{{\cal P}%
}_{\delta}+\ldots.\label{1.18}%
\end{equation}
To the $CCC\bar{\cal P}$-order, hamiltonian master equation (\ref{0.6})
yields generalized Jacobi relations,
\begin{equation}
\left(  \left\{  U_{\alpha\beta}^{\delta},T_{\gamma}\right\}  +U_{\alpha\beta
}^{\mu}U_{\mu\gamma}^{\delta}\right)  +\hbox{cycle}(\alpha,\beta,\gamma)=
U_{\alpha\beta\gamma}^{\mu\delta}T_{\mu},\label{1.19}%
\end{equation}
to determine $U_{\alpha\beta\gamma}^{\mu\delta}$.

Then, in the case of a pure rotation of constraints, Eq. (\ref{1.1})
with ``minimal'' generator (\ref{1.3}) yields the transformation law for
$U_{\alpha\beta\gamma}^{\mu\delta}$,
\[
 U_{\alpha\beta\gamma}^{\prime\delta\rho}=\left(  e^{\lambda G_{1}}\right)
_{\alpha}^{\mu}\left(  e^{\lambda G_{1}}\right)  _{\beta}^{\nu}\left(
e^{\lambda G_{1}}\right)  _{\gamma}^{\sigma}U_{\mu\nu\sigma}^{\kappa\tau
}\left(  e^{-\lambda G_{1}}\right)  _{\kappa}^{\delta}\left(  e^{-\lambda
G_{1}}\right)  _{\tau}^{\rho}+
\]
\[
+\int_{0}^{\lambda}d\lambda^{\prime}\left(  e^{(\lambda-\lambda^{\prime
})G_{1}}\right)  _{\alpha}^{\mu}\left(  e^{(\lambda-\lambda^{\prime})G_{1}%
}\right)  _{\beta}^{\nu}\left(  e^{(\lambda-\lambda^{\prime})G_{1}}\right)
_{\gamma}^{\sigma}\times
\]
\begin{equation}
 \times\left( -\left\{  G_{1\mu}^{[\kappa},U_{\nu\sigma}^{\prime\tau
]}\right\}  +\mathrm{cycle}(\mu,\nu,\sigma)\right)  \left(  e^{-(\lambda
-\lambda^{\prime})G_{1}}\right)  _{\kappa}^{\delta}\left(  e^{-(\lambda
-\lambda^{\prime})G_{1}}\right)  _{\tau}^{\rho}.\label{1.20}%
\end{equation}
By making use of (\ref{1.10}), (\ref{1.11}), one can rewrite the integral in
(\ref{1.20}) to extract all required ``transport'' terms and ``trivial shift''
ones as well. The latter have the form
\begin{equation}
A_{\alpha\beta\gamma}^{\mu\kappa\tau}T_{\mu}\left(  e^{-\lambda G_{1}%
}\right)  _{\kappa}^{\delta}\left(  e^{-\lambda G_{1}}\right)  _{\tau
}^{\rho},\label{1.21}%
\end{equation}
with $A_{\alpha\beta\gamma}^{\mu\kappa\tau}$, totally antisymmetric in
($\alpha$, $\beta$, $\gamma$) and ($\mu$, $\kappa$, $\tau$), being a
counterpart to (\ref{1.12}).

If one considers in (\ref{1.1}) more general form (\ref{1.15}) of a
generator $G$, then the expression in parentheses in the integrand in (\ref
{1.20}) extends by adding the terms
\begin{equation}
\left( -\left\{ T_{\mu }^{\prime },G_{2\nu \sigma }^{\kappa \tau
}\right\} +U_{\mu \nu }^{\prime \zeta }G_{2\zeta \sigma }^{\kappa \tau
}-G_{2\mu \nu }^{\zeta \lbrack \kappa }U_{\zeta \sigma }^{\prime \tau
]}\right) +\mathrm{cycle}(\mu ,\nu ,\sigma )+G_{3\mu \nu \sigma }^{\zeta
\kappa \tau }T_{\zeta }^{\prime }.  \label{1.22}
\end{equation}
Note that the $CCC\bar{\cal P}\bar{\cal P}\bar{\cal P}$-term in (\ref{1.15})
is the highest-order one to contribute to the transformation law for
$U_{\alpha\beta\gamma}^{\delta\rho}$.

The same as in (\ref{1.16}), by making an appropriate choice of coefficient
functions $G_{3\mu \nu \sigma }^{\zeta \kappa \tau }$ in (\ref{1.22}),
one can always compensate the trivial-shift contribution (\ref{1.21}) to
(\ref{1.20}).

\section{Quantum transformations}

So, we have shown at the classical level that pure rotations of constraints, as
defined by the generator (\ref{1.3}), yield the transformation law
(\ref{1.11}) for structure functions $U_{\alpha\beta}^{\gamma}$, with nonzero
trivial shift coefficients (\ref{1.12}) represented in terms of a rotation
matrix $G_{\alpha}^{\beta}$.

In principle, one can consider analogous transformations at the operator level
as well. An operator version of hamiltonian master equation reads
\cite{BatFra83a} - \cite{HenTei92}
\begin{equation}
\lbrack\Omega,\Omega]=0,\;\varepsilon(\Omega)=1,\;\mathrm{gh}(\Omega
)=1,\label{2.1}%
\end{equation}
where
\begin{equation}
\Omega=\Omega(q,p,C,\bar{\cal P})=\Omega^{\dagger},\label{2.2}%
\end{equation}
and the only nonzero commutator for phase variable are
\begin{equation}
\lbrack q^{j},p_{k}]=i\hbar\delta_{k}^{j},\;[C^{\alpha},\bar{ \cal P}_{\beta}]
=i\hbar\delta_{\beta}^{\alpha}.\label{2.3}%
\end{equation}
We assume the operators $q^{j}$, $p_{k}$, $C^{\alpha}$ to be hermitean, while
$\bar{\cal P}_{\beta}$ antihermitean.

One can seek for a solution for $\Omega$ in the form of a Weyl-ordered
power-series expansions in ghosts $C^{\alpha}$, $\bar{\cal P}_{\beta}$,
\cite{BatFra88}, \cite{Batt03}
\begin{equation}
\Omega=C^{\alpha}T_{\alpha}+\frac{1}{6}(C^{\beta}C^{\alpha}U_{\alpha\beta
}^{\gamma}\bar{\cal P}_{\gamma}+\bar{\cal P}_{\gamma}U_{\alpha\beta
}^{\gamma}C^{\beta}C^{\alpha}+C^{\alpha}U_{\alpha\beta}^{\gamma}
\bar{\cal P}_{\gamma}C^{\beta})+\ldots,\label{2.4}%
\end{equation}
with all coefficient operators $T_{\alpha}$, $U_{\alpha\beta}^{\gamma}$, ...
being hermitean.

To the second order in $C^{\alpha}$, Eq. (\ref{2.1}) yields the operator
involution relations \cite{BatFra88}, \cite{Batt03},
\begin{equation}
\lbrack T_{\alpha},T_{\beta}]=\frac{i\hbar}{2}(U_{\alpha\beta}^{\gamma
}T_{\gamma}+T_{\gamma}U_{\alpha\beta}^{\gamma})+\left(  \frac{i\hbar}%
{2}\right)  ^{2}[U_{\alpha\delta}^{\gamma},U_{\gamma\beta}^{\delta}%
]+\ldots.\label{2.5}%
\end{equation}

An operator version of ghost-dependent canonical transformation (\ref{1.1})
reads
\begin{equation}
i\hbar\frac{d\Omega^{\prime}}{d\lambda}=[G,\Omega^{\prime
}],\;\left.  \Omega^{\prime}\right|  _{\lambda=0}=\Omega, \label{2.6}%
\end{equation}
\begin{equation}
G=G(q,p,C,\bar{\cal P})=G^{\dagger},\;\varepsilon(G)=0,\;\hbox{gh}%
(G)=0.\label{2.6a}%
\end{equation}

One can assume a generator $G$ to be given in the form of a Weyl-ordered
power-series expansion in ghosts $C^{\alpha}$, $\bar{\cal P}_{\beta}$,
\begin{equation}
G=G_{0}+\frac{1}{2}(C^{\alpha}G_{1\alpha}^{\beta}\bar{\cal P}_{\beta
}-\bar{\cal P}_{\beta}G_{1\alpha}^{\beta}C^{\alpha})+\ldots,\label{2.9}%
\end{equation}
with all coefficient operators $G_{0}$, $G_{1\alpha}^{\beta}$, ... being
hermitean.

The transformed operator $\Omega^{\prime}$ in (\ref{2.6}) can also be sought
for in the form of a Weyl-ordered power series in ghosts,
\begin{equation}
\Omega^{\prime}=C^{\alpha}T_{\alpha}^{\prime}+\frac{1}{6}(C^{\beta}C^{\alpha
}U_{\alpha\beta}^{\prime\gamma}\bar{\cal P}_{\gamma}+\bar{\cal P}_{\gamma}
U_{\alpha\beta}^{\prime\gamma}C^{\beta}C^{\alpha}+C^{\alpha}
U_{\alpha\beta}^{\prime\gamma}\bar{\cal P}_{\gamma}C^{\beta}%
)+\ldots,\label{2.10}%
\end{equation}
with all transformed coefficient operators $T_{\alpha}^{\prime}$,
$U_{\alpha\beta}^{\prime\gamma}$, ... being hermitean.

By substituting (\ref{2.9}), (\ref{2.10}) into (\ref{2.6}), and then expanding
the latter in a Weyl-ordered power series in ghosts, one can, in principle,
derive the transformation law for all operators $T_{\alpha}$, $U_{\alpha\beta
}^{\gamma}$, ....

Thus, for example, in the case of pure rotations of constraints,
\begin{equation}
G=\frac{1}{2}(C^{\alpha}G_{1\alpha}^{\beta}\bar{\cal P}_{\beta
}-\bar{\cal P}_{\beta}G_{1\alpha}^{\beta}C^{\alpha}),\label{2.11}%
\end{equation}
Eq. (\ref{2.6}) yields to $C$- and $CC\bar{\cal P}$-orders
\begin{equation}
\frac{dT_{\alpha}^{\prime}}{d\lambda}=\frac{1}{2}(T_{\beta}^{\prime}%
G_{1\alpha}^{\beta}+G_{1\alpha}^{\beta}T_{\beta}^{\prime})+\frac{i\hbar}%
{4}[U_{\alpha\beta}^{\prime\gamma},G_{1\gamma}^{\beta}],\;\left.  T_{\alpha
}^{\prime}\right|  _{\lambda=0}=T_{\alpha},\label{2.12}%
\end{equation}
and
\[
\frac{dU_{\alpha\beta}^{\prime\gamma}}{d\lambda}=-\frac{1}%
{2}(G_{1[\alpha}^{\delta}U_{\beta]\delta}^{\prime\gamma}+
U_{\delta[\alpha}^{\prime\gamma}G_{1\beta]}^{\delta})-
\frac{1}{2}(U_{\alpha\beta}^{\prime\delta}G_{1\delta}^{\gamma}+
G_{1\delta}^{\gamma}U_{\alpha\beta}^{\prime\delta})+
\]
\begin{equation}
+(i\hbar)^{-1}[T_{[\alpha}^{\prime},G_{1\beta]}^{\gamma}]+\frac{i\hbar}%
{4}[U_{\alpha\beta\rho}^{\prime\delta\gamma},G_{\delta}^{\rho}],\;\left.
U_{\alpha\beta}^{\prime\gamma}\right|  _{\lambda=0}=U_{\alpha\beta}^{\gamma
},\;\left.  U_{\alpha\beta\varrho}^{\prime\delta\gamma}\right|  _{\lambda
=0}=U_{\alpha\beta\varrho}^{\delta\gamma},\label{2.13}%
\end{equation}
where $U_{\alpha\beta\varrho}^{\delta\gamma}$ are $CCC\bar{\cal P}
\bar{\cal P}$-order structure coefficient operators in (\ref{2.4}), and
$U_{\alpha\beta\varrho}^{\prime\delta\gamma}$ are the corresponding transformed
operators in (\ref{2.10}). These transformed operators satisfy the
$CCC\bar{\cal P}\bar{\cal P}$-order equations generated by (\ref{2.6}).

Eqs. (\ref{2.12}), (\ref{2.13}) are operator-valued counterparts to
classical Eqs. (\ref{1.4}), (\ref{1.5}). It is a characteristic feature of
(\ref{2.12}), (\ref{2.13}) and all subsequent equations generated by Eq.
(\ref{2.6}) that each of them involves the next-order structure
coefficient operators. Therefore, it is impossible to resolve (\ref{2.12})
first, and then resolve (\ref{2.13}) in turn, and so on. Instead, one faces a
difficult problem of solving an infinite chain of mutually-coupled equations.
However, these equations, when rewritten in terms of symbols w.r.t.
original phase space variables ($q,p$), can be well-resolved quasiclassically,
in terms of formal $\hbar$-power-series expansions, with explicit solution
(\ref{1.6}), (\ref{1.11}) and subsequent coefficient functions in (\ref{1.2})
being a classical approximation.

It is also worthy to mention that one can make use of a Wick basis of
constraints as represented by pairs, ($T_{\alpha}$, $T_{\alpha}^{\dagger}$),
in which case ghost sector consists of the corresponding Wick pairs,
($C^{\alpha}$, $\bar{C}_{\alpha}^{\dagger}$) and ($\bar{C}_{\alpha}$,
$C^{\dagger\alpha}$), with the only nonzero commutators
\begin{equation}
[C^{\alpha},\bar{C}_{\beta}^{\dagger}]=\delta_{\beta}^{\alpha},\quad\quad
[\bar{C}_{\alpha},C^{\dagger\beta}]=\delta_{\alpha}^{\beta},\label{2.14}%
\end{equation}
and a Wick-ordered BRST-BVF charge expands in ghost-power series in the form
\cite{BatFra88}, \cite{Batt03}
\[
\Omega=T_{\alpha}^{\dagger}C^{\alpha}+C^{\dagger\alpha}T_{\alpha}+\frac{1}%
{2}\bar{C}_{\gamma}^{\dagger}U_{\alpha\beta}^{\dagger\gamma}C^{\alpha}%
C^{\beta}+\frac{1}{2}C^{\dagger\beta}C^{\dagger\alpha}U_{\alpha\beta}^{\gamma
}\bar{C}_{\gamma}+
\]
\begin{equation}
+C^{\dagger\alpha}\bar{U}_{\alpha\beta}^{\gamma}\bar
{C}_{\gamma}C^{\beta}+C^{\dagger\beta}\bar{C}_{\gamma}^{\dagger}\bar
{U}_{\alpha\beta}^{\dagger\gamma}C^{\alpha}+\ldots.\label{2.15}%
\end{equation}

To the second order in ghosts, operator master equation (\ref{2.1}) yields the
following involution relations \cite{BatFra88}, \cite{Batt03}
\begin{equation}
[T_{\alpha},T_{\beta}]=U_{\alpha\beta}^{\gamma}T_{\gamma},\;[T_{\beta
}^{\dagger},T_{\alpha}^{\dagger}]=T_{\gamma}^{\dagger}U_{\alpha\beta}%
^{\dagger\gamma},\label{2.16}%
\end{equation}
\begin{equation}
[T_{\alpha},T_{\beta}^{\dagger}]=\bar{U}_{\alpha\beta
}^{\gamma}T_{\gamma}+T_{\gamma}^{\dagger}\bar{U}_{\beta\alpha}^{\dagger\gamma
}+\bar{U}_{\alpha\delta}^{\gamma}\bar{U}_{\beta\gamma}^{\dagger\delta}+\ldots
.\label{2.16a}%
\end{equation}
Canonical generator in (\ref{2.6}) expands in ghost power series in the form
\begin{equation}
G=G_{0}+i\hbar C^{\dagger\alpha}G_{1\alpha}^{\beta}\bar{C}_{\beta}-i\hbar
\bar{C}_{\beta}^{\dagger}G_{1\alpha}^{\dagger\beta}C^{\alpha}+\ldots
.\label{2.17}%
\end{equation}

In the case of pure rotations of constraints,
\begin{equation}
G=i\hbar C^{\dagger\alpha}G_{1\alpha}^{\beta}\bar{C}_{\beta}-i\hbar\bar
{C}_{\beta}^{\dagger}G_{1\alpha}^{\dagger\beta}C^{\alpha},\label{2.18}%
\end{equation}
we have the following transformation law for $T_{\alpha}$, $U_{\alpha\beta
}^{\gamma}$, $\bar{U}_{\alpha\beta}^{\gamma}$ and their hermite-conjugate,
\begin{equation}
\frac{dT_{\alpha}^{\prime}}{d\lambda}=G_{1\alpha}^{\beta}T_{\beta}^{\prime
},\;\left.  T_{\alpha}^{\prime}\right|  _{\lambda=0}=T_{\alpha},\label{2.19}%
\end{equation}%
\begin{equation}
\frac{dU_{\alpha\beta}^{\prime\gamma}}{d\lambda}=-G_{1[\alpha}^{\delta
}U_{\beta]\delta}^{\prime\gamma}-U_{\alpha\beta}^{\prime\delta}G_{1\delta
}^{\gamma}+[T_{[\alpha}^{\prime},G_{1\beta]}^{\gamma}],\;\left.
U_{\alpha\beta}^{\prime\gamma}\right|  _{\lambda=0}=U_{\alpha\beta}^{\gamma
},\label{2.20}%
\end{equation}%
\begin{equation}
\frac{d\bar{U}_{\alpha\beta}^{\prime\gamma}}{d\lambda}=G_{1\alpha}^{\delta
}\bar{U}_{\delta\beta}^{\prime\gamma}+\bar{U}_{\alpha\delta}^{\prime\gamma
}G_{1\beta}^{\dagger\delta}-\bar{U}_{\alpha\beta}^{\prime\delta}G_{1\delta
}^{\gamma}+[G_{1\alpha}^{\gamma}T_{\beta}^{\dagger\prime},],\;\left.  \bar
{U}_{\alpha\beta}^{\prime\gamma}\right|  _{\lambda=0}=\bar{U}_{\alpha\beta
}^{\gamma}.\label{2.21}%
\end{equation}
Eq. (\ref{2.19}) can be formally resolved itself. Then, in turn, Eqs.
(\ref{2.20}), (\ref{2.21}) can be formally resolved as well. Thus we obtain the
following solution
\begin{equation}
T_{\alpha }^{\prime }=\left( e^{\lambda G_{1}}\right) _{\alpha }^{\beta
}T_{\beta },  \label{2.22}
\end{equation}
\begin{equation}
U_{\alpha \beta }^{\prime \gamma }=\left( e^{\lambda F}\right) _{\alpha
\beta }^{\mu \nu }U_{\mu \nu }^{\rho }\left( e^{-\lambda G_{1}}\right)
_{\rho }^{\gamma }+\int_{0}^{\lambda }d\lambda ^{\prime }\left( e^{(\lambda
-\lambda ^{\prime })F}\right) _{\alpha \beta }^{\mu \nu }[T_{[\mu }^{\prime
},G_{1\nu ]}^{\rho }]\left( e^{-(\lambda -\lambda ^{\prime })G_{1}}\right)
_{\rho }^{\gamma },  \label{2.23}
\end{equation}
\begin{equation}
\bar{U}_{\alpha \beta }^{\prime \gamma }=\left( e^{\lambda G_{1}}\right)
_{\alpha }^{\mu }\bar{U}_{\mu \rho }^{\delta }\left( e^{\lambda \bar{F}%
}\right) _{\delta \beta }^{\rho \gamma }+\int_{0}^{\lambda }d\lambda
^{\prime }\left( e^{(\lambda -\lambda ^{\prime })G_{1}}\right) _{\alpha
}^{\mu }[G_{1\mu }^{\delta },T_{\rho }^{\dagger \prime }]\left( e^{(\lambda
-\lambda ^{\prime })\bar{F}}\right) _{\delta \beta }^{\rho \gamma },
\label{2.24}
\end{equation}
where
\begin{equation}
F_{\alpha \beta }^{\mu \nu }=\frac{1}{2}G_{1[\alpha }^{[\mu }\delta _{\beta
]}^{\nu ]},\;\bar{F}_{\alpha \beta }^{\mu \nu }=\delta _{\alpha }^{\nu
}G_{1\beta }^{\dagger \mu }-\delta _{\beta }^{\mu }G_{1\alpha }^{\nu }.
\label{2.25}
\end{equation}
In Eqs. (\ref{2.23}), (\ref{2.24}), the exponentials $e^{\lambda F}$, $%
e^{\lambda \bar{F}}$ are defined as follows
\begin{equation}
\frac{d}{d\lambda }\left( e^{\lambda F}\right) _{\alpha \beta }^{\mu \nu
}=F_{\alpha \beta }^{\gamma \delta }\left( e^{\lambda F}\right) _{\gamma
\delta }^{\mu \nu },\;\left. \left( e^{\lambda F}\right) _{\alpha \beta
}^{\mu \nu }\right| _{\lambda =0}=\frac{1}{2}\delta _{\alpha }^{[\mu }\delta
_{\beta }^{\nu ]},  \label{2.26}
\end{equation}
\begin{equation}
\frac{d}{d\lambda }\left( e^{\lambda \bar{F}}\right) _{\alpha \beta }^{\mu
\nu }=\left( e^{\lambda \bar{F}}\right) _{\alpha \gamma }^{\mu \delta }\bar{F%
}_{\delta \beta }^{\gamma \nu },\;\left. \left( e^{\lambda \bar{F}}\right)
_{\alpha \beta }^{\mu \nu }\right| _{\lambda =0}=\delta _{\alpha }^{\nu
}\delta _{\beta }^{\mu }.  \label{2.27}
\end{equation}
These equations mean that $e^{\lambda F}$ and $e^{\lambda
\bar{F}}$ are the standard exponential $F$- and $\bar{F}$-power
series expansions, in which operator-valued matrix $F_{\alpha
\beta }^{\mu \nu }$ is labelled with
``left'' and ``right'' indices being antisymmetric sectors of the sets ($%
\alpha $, $\beta $) and ($\mu $, $\nu $), respectively, while $\bar{F}%
_{\alpha \beta }^{\mu \nu }$ is labelled with ``left'' and
``right'' indices being the sets ($\alpha $, $\mu $) and ($\nu $,
$\beta $), respectively.

Transformations (\ref{2.23}), (\ref{2.24}) can be rewritten in the form
which allows for explicit extraction of the trivial shift part as defined by
our general rule: this part should yield no contribution to the involution
relations; on the other hand, it should be removable by the effect of a
fourth-order contribution to the ghost-dependent canonical transformation. It
is obvious that this fourth-order contribution does not change the constraint
operators themselves. We have
\[
U_{\alpha \beta }^{\prime \gamma }=\frac{1}{2}
\left(e^{\lambda G_1}\right)_{[\alpha}^{\mu}\left(e^{\lambda G_1}
\right)^\nu_{\beta]}U_{\mu \nu }^{\rho }\left( e^{-\lambda G_{1}}\right)
_{\rho }^{\gamma }+\left( e^{\lambda G_{1}}\right) _{[\alpha }^{\mu }[T_{\mu
},\left( e^{\lambda G_{1}}\right) _{\beta ]}^{\nu }]\left( e^{-\lambda
G_{1}}\right) _{\nu }^{\gamma }+
\]
\begin{equation}
+\frac{1}{2}[\left( e^{\lambda G_{1}}\right) _{\alpha }^{\{\mu },\left(
e^{\lambda G_{1}}\right) _{\beta }^{\nu \}}]T_\mu
\left( e^{-\lambda G_{1}}\right)
_{\nu }^{\gamma }+\frac{1}{2}A_{\alpha \beta }^{\mu \nu }(T_{[\mu }\delta
_{\nu ]}^{\rho }-U_{\mu \nu }^{\rho })\left( e^{-\lambda G_{1}}\right)
_{\rho }^{\gamma },  \label{2.28}
\end{equation}
where
\begin{equation}
A_{\alpha \beta }^{\mu \nu }=\frac{1}{2}\int_{0}^{\lambda }d\lambda ^{\prime
}\left( e^{(\lambda -\lambda ^{\prime })F}\right) _{\alpha \beta }^{\rho
\sigma }[\left( e^{\lambda ^{\prime }G_{1}}\right) _{[\rho }^{[\mu
},G_{1\sigma ]}^{\tau }]\left( e^{\lambda ^{\prime }G_{1}}\right) _{\tau
}^{\nu ]}  \label{2.29}
\end{equation}
In Eq. (\ref{2.28}), there is an ambiguity in operators
$A_{\alpha\beta}^{\mu\nu}$ caused by their own trivial shift,
\begin{equation}
A_{\alpha\beta}^{\mu\nu}\;\rightarrow A_{\alpha\beta}^{\mu\nu}+
\frac{1}{6}A_{\alpha\beta}^{\zeta\kappa\sigma}\left[(T_\zeta
\delta_\kappa^{[\mu}\delta_\sigma^{\nu]}-U_{\zeta\kappa}^{[\mu}
\delta_\sigma^{\nu]})+\hbox{cycle}(\zeta,\kappa,\sigma)-
U_{\zeta\kappa\sigma}^{\mu\nu}\right],  \label{2.29a}
\end{equation}
with arbitrary operator coefficients $A_{\alpha\beta}^{\zeta\kappa\sigma}$
totally antisymmetric in ($\alpha$, $\beta$), ($\zeta$, $\kappa$, $\sigma$),
and $U_{\zeta\kappa\sigma}^{\mu\nu}$ totally antisymmetric in ($\zeta$,
$\kappa$, $\sigma$), ($\mu$, $\nu$), being determined, up to their own
ambiguity, by quantum counterpart of the relation (\ref{1.19}),
\begin{equation}
\left([U_{\alpha\beta}^\delta,T_\gamma]+U_{\alpha\beta}^\mu
U_{\mu\gamma}^\delta\right)+\hbox{cycle}(\alpha,\beta,\gamma)=
\frac{1}{2}U_{\alpha\beta\gamma}^{\mu\nu}(T_{[\mu }\delta_{\nu]}^\delta-
U_{\mu\nu}^\delta).  \label{2.29b}
\end{equation}
In fact, these $U_{\alpha\beta\gamma}^{\mu\nu}$ are coefficients in
$C^\dagger C^\dagger C^\dagger\bar{C}\bar{C}$-contribution to the      operator
(\ref{2.15}).

Expressions (\ref{2.28}), (\ref{2.29}) are exact quantum counterparts of the
corresponding classical expressions (\ref{1.11}), (\ref{1.12}), which
correspondence shows that classical and quantum descriptions are given by
universal formulae. In Eq. (\ref{2.28}), the first, second and third terms
in\ its r.h.s. represent the ``geometric'' part of the transformation, which
is actually required by the covariance of the involution relations, while
the fourth term represents the trivial-shift part. It is very instructive to
note that the integrand in (\ref{2.29}) is a total $\lambda ^{\prime }$%
-derivative so that one can take the integral explicitly to get

\begin{equation}
A_{\alpha \beta }^{\mu \nu }=\frac{1}{2}\left( e^{\lambda G_{1}}\right)
_{[\alpha }^{[\mu }\left( e^{\lambda G_{1}}\right) _{\beta ]}^{\nu
]}-2\left( e^{\lambda F}\right) _{\alpha \beta }^{\mu \nu }.  \label{2.30}
\end{equation}

Analogous consideration can be performed for
$\bar{U}_{\alpha\beta}^{\prime\gamma}$ and all higher-order structure
coefficient operators.

\section{Conclusion}

In the above sections, we have studied, at the classical and quantum level,
the transformation law for structure coefficients of hamiltonian gauge
algebra under rotations of constraints. We have shown that the
transformation law splits naturally into two parts called ``geometric'' and
``trivial-shift'', respectively. Geometric part is actually required by the
covariance of the corresponding structure relations, while trivial-shift
part yields no contribution to the structure relations, and can always be
removed by ghost-depending canonical transformation which does not change
the constraints.

Further details and some new aspects of the transformation properties of
hamiltonian gauge algebra will be considered elsewhere.

\textbf{Acknowledgements}

The authors are grateful to Maxim Grigoriev and Alexei Semikhatov for
fruitful discussions.  The work of I.A.  Batalin is supported in part by the
grant LSS-1578.2003.2, the RFBR grant 02-01-00930, and the
INTAS grant 00-00262.  The work of I.V. Tyutin is supported in part by the
grant LSS-1578.2003.2, the RFBR grant 02-02-16944, and the
INTAS grant 00-00262.

\def\theequation{A.\arabic{equation}}
\setcounter{equation}{0}

\section{Appendix}

Here, by making an appropriate choice of functions $G_{1\alpha}^\beta$, we
show explicitly that $(A_{\alpha\beta}^{\mu\nu})_3\not\equiv O(T)$, where
$(A_{\alpha\beta}^{\mu\nu})_3$ are given by Eq. (\ref{1.14}).

Let the index $\alpha$ of the constraints $T_\alpha$ be split into two subsets
$\alpha_1$, $\alpha_2$, so that $\alpha =(\alpha _1,\alpha_2)$, and the matrix
of functions $G_{1\alpha }^{\beta }$ is chosen in the form
\begin{equation}
G_{1\alpha }^{\beta }=\left(
\begin{array}{cc}
0 & M_{\alpha _{1}}^{\beta _{2}} \\
N_{\alpha _{2}}^{\beta _{1}} & 0
\end{array}
\right) .  \label{A.1}
\end{equation}
Then we have
\begin{equation}
\left( A_{\alpha _{1}\beta _{1}}^{\mu _{2}\nu _{1}}\right) _{3}=\frac{%
\lambda ^{3}}{6}\left( \left\{ M_{\alpha _{1}}^{\mu _{2}},M_{\beta
_{1}}^{\rho _{2}}\right\} N_{\rho _{2}}^{\nu _{1}}-\left\{ M_{\beta
_{1}}^{\mu _{2}},M_{\alpha _{1}}^{\rho _{2}}\right\} N_{\rho _{2}}^{\nu
_{1}}\right) .  \label{A.2}
\end{equation}
Further, let us assume that the indices ($\alpha _{1}$, $\beta _{1},\nu _{1})
$ and $\mu _{2}$ are split, in turn, into two subsets, $\alpha _{1}=(\alpha
_{11},\alpha _{12})$, $\mu _{2}=(\mu _{21},\mu _{22})$, and the matrices of
functions $M_{\alpha _{1}}^{\mu _{2}}$ and $N_{\rho _{2}}^{\nu _{1}}$ are
chosen in the form
\begin{equation}
M_{\alpha _{1}}^{\mu _{2}}=\left(
\begin{array}{cc}
0 & a_{\alpha _{11}}^{\mu _{22}} \\
b_{\alpha _{12}}^{\mu _{21}} & 0
\end{array}
\right) ,\;N_{\rho _{2}}^{\nu _{1}}=\left(
\begin{array}{cc}
0 & c_{\rho _{21}}^{\nu _{12}} \\
d_{\rho _{22}}^{\nu _{11}} & 0
\end{array}
\right) ,  \label{A.3}
\end{equation}
with the block $c_{\rho _{21}}^{\nu _{12}}$ being invertible. Then, it
follows that
\begin{equation}
\left( A_{\alpha _{11}\beta _{12}}^{\mu _{22}\nu _{12}}\right) _{3}=\frac{%
\lambda ^{3}}{6}\left\{ a_{\alpha _{11}}^{\mu _{22}},b_{\beta _{12}}^{\rho
_{21}}\right\} c_{\rho _{21}}^{\nu _{12}}.  \label{A.4}
\end{equation}
It is obvious that there exist matrices $a_{\alpha _{11}}^{\mu _{22}}$ and $%
b_{\beta _{12}}^{\rho _{21}}$ such that $\left\{ a_{\alpha _{11}}^{\mu
_{22}},b_{\beta _{12}}^{\rho _{21}}\right\} \not\equiv $ $O(T)$.

\bigskip\

\end{document}